\documentclass[draft]{agujournal2019}
\usepackage{url} 
\usepackage{soul}
\usepackage{amssymb}
\usepackage{subcaption}

\draftfalse

\journalname{Space Weather}
\hypersetup{final}
\newcounter{magicrownumbers}

\begin{document}

%
%


\title{Beacon2Science: Enhancing STEREO/HI beacon data with machine learning for efficient CME tracking}

%
%




\authors{J. Le Lou\"edec\affil{1}, M. Bauer\affil{1,2}, T. Amerstorfer\affil{1}, J.A. Davies\affil{3}}





\affiliation{1}{Austrian Space Weather Office, GeoSphere Austria, Reininghausstraße 3, Graz, 8020, Austria}
\affiliation{2}{Institute of Physics, University of Graz, Universit\"atsplatz 5, Graz, 8010, Austria}
\affiliation{3}{RAL Space, STFC Rutherford Appleton Laboratory, Didcot, UK}

\correspondingauthor{Justin Le Lou\"edec}{justin.lelouedec@geosphere.at}

\begin{keypoints}
\item	Near-real-time CME arrival time prediction with HI data is challenging due to the low resolution and high noise of beacon data.
\item	We propose the new ``Beacon2Science'' machine-learning pipeline to upsample, denoise, and increase the temporal resolution of beacon data.
\item	The improved images show better CME visibility and better tracking of the CME leading edge tracks closer to those made with science data.
\end{keypoints}

\begin{abstract}
Observing and forecasting coronal mass ejections (CME) in real-time is crucial due to the strong geomagnetic storms they can generate that can have a potentially damaging effect, for example, on satellites and electrical devices. With its near-real-time availability, STEREO/HI beacon data is the perfect candidate for early forecasting of CMEs. However, previous work concluded that CME arrival prediction based on beacon data could not achieve the same accuracy as with high-resolution science data due to data gaps and lower quality. We present our novel machine-learning pipeline entitled ``Beacon2Science'', bridging the gap between beacon and science data to improve CME tracking. Through this pipeline, we first enhance the quality (signal-to-noise ratio and spatial resolution) of beacon data. We then increase the time resolution of enhanced beacon images through learned interpolation to match science data's 40-minute resolution. We maximize information coherence between consecutive frames with adapted model architecture and loss functions through the different steps.
The improved beacon images are comparable to science data, showing better CME visibility than the original beacon data. Furthermore, we compare CMEs tracked in beacon, enhanced beacon, and science images. The tracks extracted from enhanced beacon data are closer to those from science images, with a mean average error of $\sim 0.5 ^\circ$ of elongation compared to $1^\circ$ with original beacon data. The work presented in this paper paves the way for its application to forthcoming missions such as Vigil and PUNCH.

\end{abstract}

\section*{Plain Language Summary}
Coronal mass ejections (CMEs) are large eruptions of plasma from the Sun, which can disrupt satellites, power grids, and communication systems when reaching Earth. To better study and predict their arrival, the STEREO mission was sent to capture pictures of CMEs in visible light using a heliospheric imager. 
These images are transmitted to Earth at high quality with a delay of 3 to 4 days (so-called science data, which is not helpful for real-time prediction of a CME),  or at very low quality in near-real-time called beacon data. To use this beacon data, we use novel machine-learning techniques to increase the spatial and temporal resolution, reduce the noise, and create intermediate images for smoother visualization. With improved data quality and visibility of CMEs, we track them through space toward Earth nearly as accurately as with science data. 
This research is important for studying CMEs, as it can be applied to improve real-time predictions of their arrival at Earth. Furthermore, we can apply our method to future space weather missions. One of them is Vigil, a planned ESA satellite that will monitor CMEs from a unique position in space (L5) and help to protect human technology and infrastructure on Earth.

\section{Introduction}


Coronal mass ejections (CMEs) are plasma and magnetic flux eruptions from the Sun that propagate outward through the heliosphere. The interaction of the fast-expelled material with planets and their magnetospheres leads to various phenomena that drive most space weather effects. While these events can be harmless, more ``geo-effective'' events can induce intense and destructive geomagnetic storms with substantial impacts, such as satellite orbit changes, electronic device perturbations on the planet's surface and radio transmission disturbances. Such impacts are getting even more critical with the increasing reliance on technology and continuous power supply to various infrastructures, thus increasing the importance of real-time CME predictions \cite{feynman2000space,cannon2013,Eastwood2018QuantifyingTE,Oughton2018ARA,Riley2018ExtremeSW}.

Coronagraph data, i.e.\ white-light images from around the Sun up to 30 solar radii, are the primary data source for modeling the evolution of CMEs. However, arrival predictions based on CME characteristics extracted from these data often come with high uncertainties due to the imprecise initial parameters \cite{sin22,ver23}, even more so when limited numbers of vantage points are used as described by \citeA{palkay24}.

Even though coronagraph data is essential in operational space weather forecasting as it provides early initial CME parameters, integrating additional data, such as from heliospheric imagers (HI), particularly when they provide a side-view on CMEs, is essential for improving forecasting accuracy and mitigating uncertainties.

For much of the time since their launch at the end of 2006, HI1 and HI2~\cite{eyl09} onboard the Solar TErrestrial RElations Observatory \cite<STEREO;>{kai08} have continuously monitored heliocentric distances between 12--215 solar radii (line-of-sight). STEREO provides science-grade data with a few days' latency and near-real-time beacon data with lower resolution and signal-to-noise-ratio. STEREO consists of twin-spacecraft STEREO-A(head) and STEREO-B(ehind), with STEREO-A and STEREO-B orbiting in an Earth-like orbit around the Sun at around 1 AU (STEREO-B with an orbit slightly larger than 1 AU such that it lags behind Earth and STEREO-A with a slightly smaller orbit than 1 AU such that it leads Earth). This configuration attains an angular separation to allow for observing space weather phenomena from two distinct vantage points. Contact with STEREO-B was lost in 2014 and data was unavailable/highly limited from STEREO-A for about a year around superior conjunction. But STEREO-A offers close to 18 years of observations of Earth-directed space weather events.

HI cameras are also found onboard of Parker Solar Probe \cite<PSP;>{fox16} with the Wide-Field Imager for Solar Probe \cite<WISPR;>{vou16} and on Solar Orbiter \cite<SolO;>{mue20} with the Solar Orbiter Heliospheric Imager \cite<SolOHI;>{how20}. However, these are only available for specific events and periods and provide no real-time data. The recently launched Polarimeter to UNify the Corona and Heliosphere (PUNCH, \cite{deforest2022polarimeter}) mission carries both a narrow field imager (NFI) and a wide field imager (WFI) onboard, and will provide polarized HI in near-real-time.

HI data has been utilized to develop forecasting models, notably the ELlipse Evolution model based on HI \cite<ELEvoHI;>{rol16, ame18}, which can be run such that it relies solely on the utilization of HI data without the necessity of integrating CME parameters measured from coronagraph data.
However, one of the main current challenges with integrating heliospheric imager data for CME arrival time and speed prediction is the absence of acceptable quality data in real-time \cite{tuc15,bau21}. 
While STEREO/HI data is used to improve the forecasting accuracy for well-observed CMEs using its science data, a posteriori, the beacon mode, due to low broadcast priority, suffers from four main drawbacks: lower spatial resolution, lower temporal resolution, significant data gaps, and for HI1, the usage of lossy compression.

Future missions, such as Vigil, planned for launch in 2031, aim to provide continuous observations between the Sun and Earth from the fifth Lagrangian point (L5) of the Sun-Earth system, $60^\circ$ behind Earth in its orbit. This side-view onto Earth-directed CMEs will provide data optimized to space weather forecasting and the mission should deliver adequate time and spatial resolution in real-time.

In the meantime, as we approach the maximum of solar cycle 25, we propose to leverage the STEREO mission's data heritage to develop and validate a novel pipeline meant to alleviate some of the challenges with the current STEREO beacon data. With STEREO-A approaching the fourth Lagrangian point (L4), our pipeline will allow to use STEREO-A/HI real-time data now and validate models for CME predictions to prepare for future missions such as Vigil.

Improving data quality during postprocessing is key to balancing the challenges of capturing and transmitting high-quality data in near-real-time for faster forecasting with more usable data. \citeA{deforest2017noise} proposed a noise-gating technique using sequences of images to improve solar and heliospheric imaging. The method, however, struggles with weak signal features, a common occurrence in HI imaging of CMEs. \citeA{jarolim2020image}, introduced a machine learning method for providing a reliable image-quality assessment of the solar disk in real-time based on Generative Adversarial Networks~\cite<GANs>{goodfellow2014generative}, showing use cases where data-driven approaches can leverage different data quality levels to extract noise and other perturbations. More recently, \citeA{schirninger2025deep} employed neural networks to combine stacks of low-resolution observations into high-resolution and quality representations to allow the observation at large scales of the solar atmosphere.

In this work, we present a method to enhance the quality and usability of HI beacon data for CME detection and tracking. Our machine learning-based method significantly reduces noise in beacon data while preserving CME structures, thus improving the visibility of key features. Additionally, we introduce an interpolation technique that enhances temporal resolution while maintaining the consistency of the CME structure, making it easier to track CME evolution across consecutive frames. Furthermore, we integrate these enhancements into a new tracking tool that compliments traditional tracking of CME features in time-elongation maps \cite<J-maps;>{she99,dav09b} with the additional spatial context of the full images. 
We thoroughly evaluate the extent of the improvement in the quality of beacon data and track CMEs across different data types. Comparing CME tracks allows us to assess the full extent of our method's improvement on the CME visibility and its impact.

\section{Data}
As part of the SECCHI instrument suite, the wide-angle HI data collected by the Solar TErrestrial RElations Observatory \cite<STEREO;>{kai08,Biesecker2008} offers an important data set to assess the future application of space weather forecasting facilities for Vigil. Situated at a heliocentric distance of roughly one AU, STEREO-A offers close to 18 years of data and crossed the Lagrangian points 4 and 5, which are considered to be ideal observation points for Earth-directed transient events. The mission provides two data transmission types, with a low-resolution feed called beacon mode providing HI data at a 120-minute cadence in near-real-time and a mode transmitting high-resolution data with a latency of a few days. In this work, we focus on HI1 data as the spatial and temporal resolution are the most reduced between beacon and science data, due to ICER lossy compression and cadenced reduction from 40-minutes to 120-minutes.
This section introduces the beacon data type, its processing, and the creation of the datasets used in this paper.

\subsection{The STEREO Beacon Mode}

\begin{figure}[t]
        \centering
        \includegraphics[width=\linewidth]{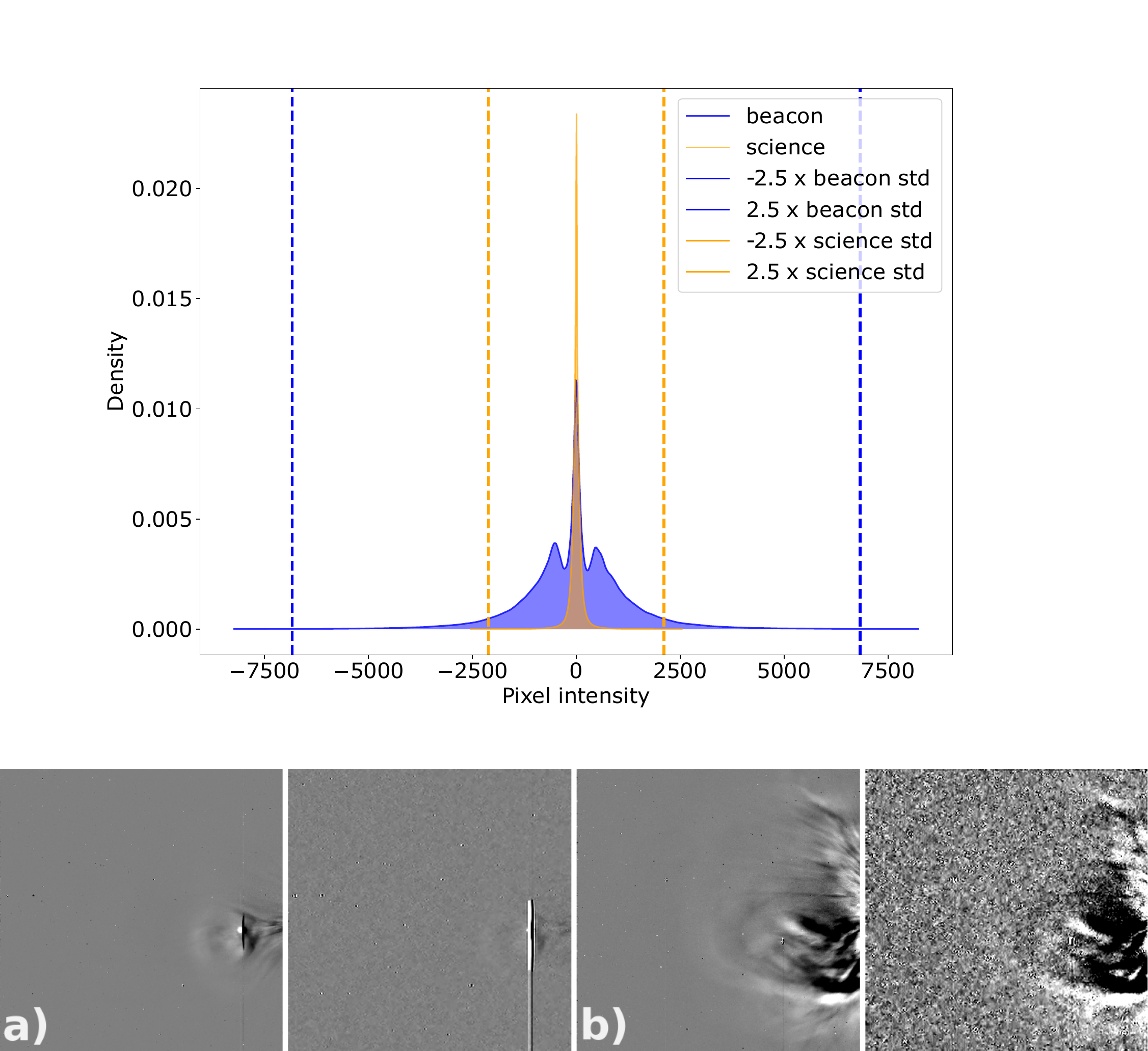} 
        \caption{(Top): Comparison of the average pixel intensity of beacon compared to science running difference images. We indicate the $\beta$ choice of 2.5 for normalization. (Bottom): Comparison of two science and beacon pairs (a and b). Each science running difference is on the left, and the beacon is on the right.}
        \label{fig:beaconvsscience}
\end{figure}

HI beacon mode data is a subset of the complete science data, compressed and binned before being broadcast to Earth. There, it is received by a small set of ground stations as a near-real-time product (based on science images) and by the Deep Space Network (DSN) together with science data a few days later. The near-real-time data based on science images selected at a 120-minute interval instead of the 40-minute cadence captured on board. During optimal broadcast and reception conditions, one in every three HI1 science images is being broadcast to Earth in the beacon data stream. 
Due to the limited broadcast rate available for the beacon mode, high compression followed by binning and, thus, degradation of the science data is necessary to offer near-real-time HI1 beacon data. We show in Figure~\ref{fig:beaconvsscience} examples of beacon and science data. Even though some visual similarities are observed, the compression of beacon data increases the noise and decreases the visibility and details (due to pixel binning) of important CME features, such as the front as it gets away from the Sun. 
Increasing the spatial resolution of beacon data is only part of the problem, and improving the signal-to-noise ratio is of higher importance for better visibility of CMEs and separating their fronts from the background noise as far as possible throughout the heliosphere, leading to better arrival forecasting.

\subsection{Data Reduction and Processing}

Following \citeA{bau21}, we use the open-source Python implementation of the IDL SolarSoft \texttt{secchi\_prep.pro} routine to process Level 0.5 HI1 beacon and science images into Level 1 data (See Section~\ref{sec:code} for details).

This procedure includes trimming the over- and underscan regions, correcting for on-board operations involving pixel intensities that are not related to compression, and applying standard astrophysical data reduction techniques, as outlined in \citeA{eyl09}. These standard procedures involve bias subtraction, removal of saturated CCD columns, flat-field correction, and replacing cosmic ray pixel counts with values from surrounding pixels values. For STEREO/HI data, it is necessary to correct for smearing effects caused by the shutter-less readout. The image is then converted to units of S10, where one S10 unit represents the flux of a tenth magnitude star of solar spectral type, distributed over 1 square degree of sky. To ensure precise alignment, the spacecraft’s pointing information is updated in the science file’s header using the most recent star-fitting parameters. However, for beacon data, a standard fixed offset is applied. A more detailed overview of the STEREO/HI pointing calibration process can be found in \citeA{bro09} and \citeA{tap22}.

The Level 2 data used in this work is derived from Level 1 data by removing the background signal, primarily composed of the F-corona, to enhance the visibility of CMEs. This is achieved by computing the pixel-wise median of images from the preceding five days and subtracting it from the current image.

Once Level 2 data is obtained, we compute running differences between pairs of images. As CMEs and other transient objects become fainter as they move away from the Sun and traverse the HI1 field of view, running differences are commonly used to enhance visibility and facilitate analysis. However, the use of running difference schemes brings with it the possibility of introducing artifacts into a given image, stemming from anomalous signals in the previous image. Despite this trade-off, we consider that the significant improvement in CME visibility justifies the use of running difference imaging in the context of this work. The time interval between the subsequent images used to create running differences, referred to as $\Delta_t$, is 120 minutes for beacon data and 40 minutes for science data (as displayed in Figure~\ref{fig:beaconvsscience}). This is the minimum possible as it corresponds to the native data cadence in each case. This difference leads to less detailed features in beacon running differences, but the leading edge and core of the CMEs should still be found at the same elongation.

To use the running differences as part of the machine learning pipeline and have more stable and efficient training, we need to normalize their values in the $[0,1]$ range. Usually, to visualize running differences, minimum and maximum values used for normalization are handpicked for each image. We propose choosing a fixed factor $\beta$ of the standard deviation of values in each image to process automatically all the data in our dataset. In Figure~\ref{fig:beaconvsscience}, we show the average histogram of beacon and science running differences for $\approx 2800$ test running differences. Science has higher maximum and minimum values, but the distribution spread is smaller than beacon. We use $\beta = 2.5$ times the image's standard deviation of pixel values as illustrated in Figure~\ref{fig:beaconvsscience}, to normalize beacon and science images without losing important information.

\subsection{Dataset Creation}

Ultimately, the goal of the method introduced in this paper is to provide real-time enhancement of STEREO/HI data for CME analysis and forecasting their arrival at a certain target, most importantly at Earth. Thus, we aim to create a large training dataset comprising a maximum of CME events captured by the STEREO/HI mission. To not overextend the scope of the paper, we focus on STEREO-A and the HI1 instrument. For this purpose, we use the HELCATS WP2 HICAT catalog~\cite{Helcats2018} and all its CME events until May 2023 observed by STEREO-A (1236 events). To cover the passage of a CME through the entire HI1 field of view, we download three days for each event (beacon and corresponding science images), starting with the day when the CME appears, while removing days that are duplicated for different CMEs. Such data coverage also allows us to have a good proportion of images without a CME visible to avoid a bias toward recreating CMEs in the enhanced images. The training set contains 21234 pairs of beacon/science running difference images.

To create a benchmark test set, we curate a set of Earth-directed CME events observed within the STEREO/HI field of view to evaluate our method on a dataset with enough variation and complexity. We use the HELCATS WP3 HIGeoCAT catalog \cite{barnes2019cmes} to filter events of interest. We use the following criteria :
\begin{enumerate}
    \item CME apex direction within $\pm 30^\circ$ heliospheric longitude \cite<Heliocentric Earth Equatorial, HEEQ;>{thompson2006coordinate}.
    \item Five beacon images during the first day of the event.
    \item Imaged by STEREO-A
    \item CME observed within the ranges [2008--2012] and [2018--2022] to correspond to observations taken from around L4 and L5.
\end{enumerate}
With such criteria, we obtain 295 events; we further narrow the selection to 48 events, with various degrees of complexity and visibility, to better assess our algorithm on challenging and simpler examples. 

For each event, we use the available data three days before and after the CME's first appearance in the HI1 field of view (for enough data to validate our model), which equates to seven days of images (beacon and corresponding science images). Including these extra days ensures enough data to evaluate the quality of the generated images in case of significant data gaps and to include the whole CME tracks in case of slow events.

\begin{figure}
\centering
    \includegraphics[width=\linewidth]{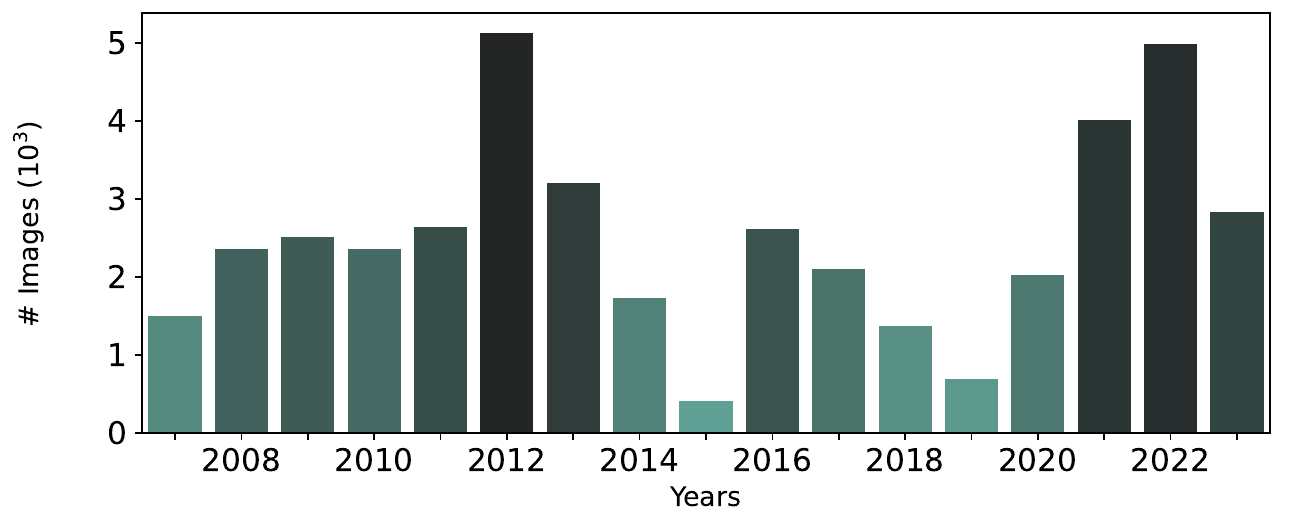}
    \caption{Summary of the training dataset, with the number of images obtained yearly from the HELCATS WP2 events}
    \label{fig:dataset1}
\end{figure}

\begin{figure}
\centering
\includegraphics[width=\linewidth]{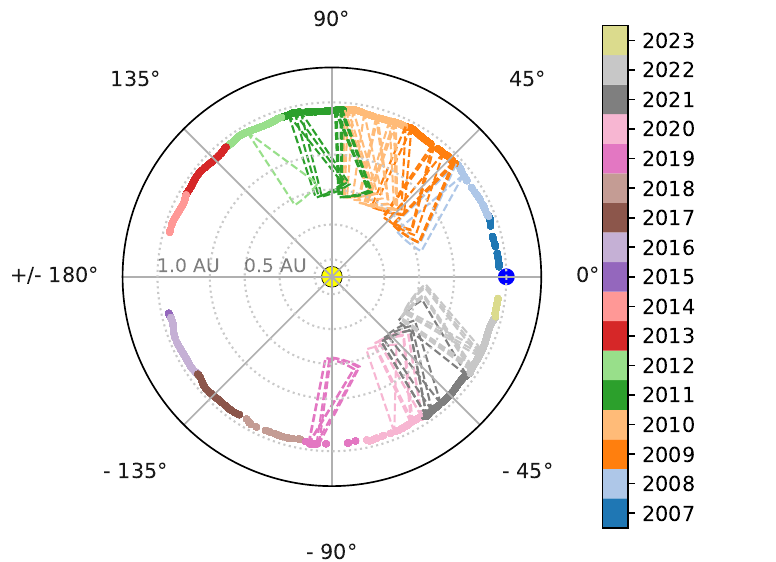}
    \caption{The STEREO-A position in HEE coordinates for training set (dots), with the HI1 field-of-view of the selected test events (cones).}
    \label{fig:dataset2}
\end{figure}

Finally, we remove all the days related to our test set's events from the training set to avoid data leakage.
We introduce our training dataset in Figure~\ref{fig:dataset1} with the number of images per year, and in Figure~\ref{fig:dataset2} the position of the spacecraft \cite<in Heliocentric Earth Ecliptic coordinates, HEE;>{thompson2006coordinate} for our training and test set combined. This curated test set includes 2176 beacon/science pairs of images.

\section{Beacon2Science Pipeline}

Beacon data from STEREO/HI1 suffers from two shortcomings: (1) the heavy compression combined with the binning of the images and (2) very low cadence compared to the science data equivalent. We aim to improve these aspects through two different neural networks forming the Beacon2Science pipeline, with the output of the first neural network called E-beacon (enhanced beacon) used as input to the second neural network producing IE-beacon (interpolated enhanced beacon).
Finally, we introduce a novel tool (see Section~\ref{sec:tool}) for directly tracking CMEs and other features of interest within the STEREO/HI1 images combined with J-maps for better identification and tracking.

\subsection{Improving Beacon Resolution and Signal-to-noise Ratio}\label{seq:NN1}

To transform near-real-time beacon data into enhanced E-beacon frames, we use a data-driven approach based on GAN. A GAN consists of two neural networks: a generator that creates new images, and a discriminator that tries to distinguish between real and generated images. These two networks are trained together in competition, encouraging the generator to produce outputs that are both realistic and structurally accurate. The end goal is for the generator to create images that so closely resemble the real data that the discriminator is unable to tell them apart. After training is done, the discriminator is no longer needed.

In our case, the generator (G) takes a beacon running-difference frame as input and outputs a corresponding E-beacon frame. At the same time, the discriminator (D) is trained to decide whether a given image is either a true science frame or an E-beacon frame produced by the generator.

Our approach is inspired by pix2pix\cite{isola2017image}, a GAN framework used to translate images from one type to another (e.g., grayscale to color, noisy to clean). To further improve resolution, we draw on techniques from ESRGAN (Enhanced Super-Resolution Generative Adversarial Networks)\cite{wang2018esrgan}, a super-resolution method. In particular, we use PixelShuffle layers~\cite{shi2016real} at the end of the generator. These layers employ a novel rearrangement technique, combined with learned weights to increase the spatial resolution of the input image, increasing its size by a factor of $X$ ($2\times$ or $4\times$ the input resolution).

\begin{figure}[h!]
    \centering
    \includegraphics[width=\textwidth]{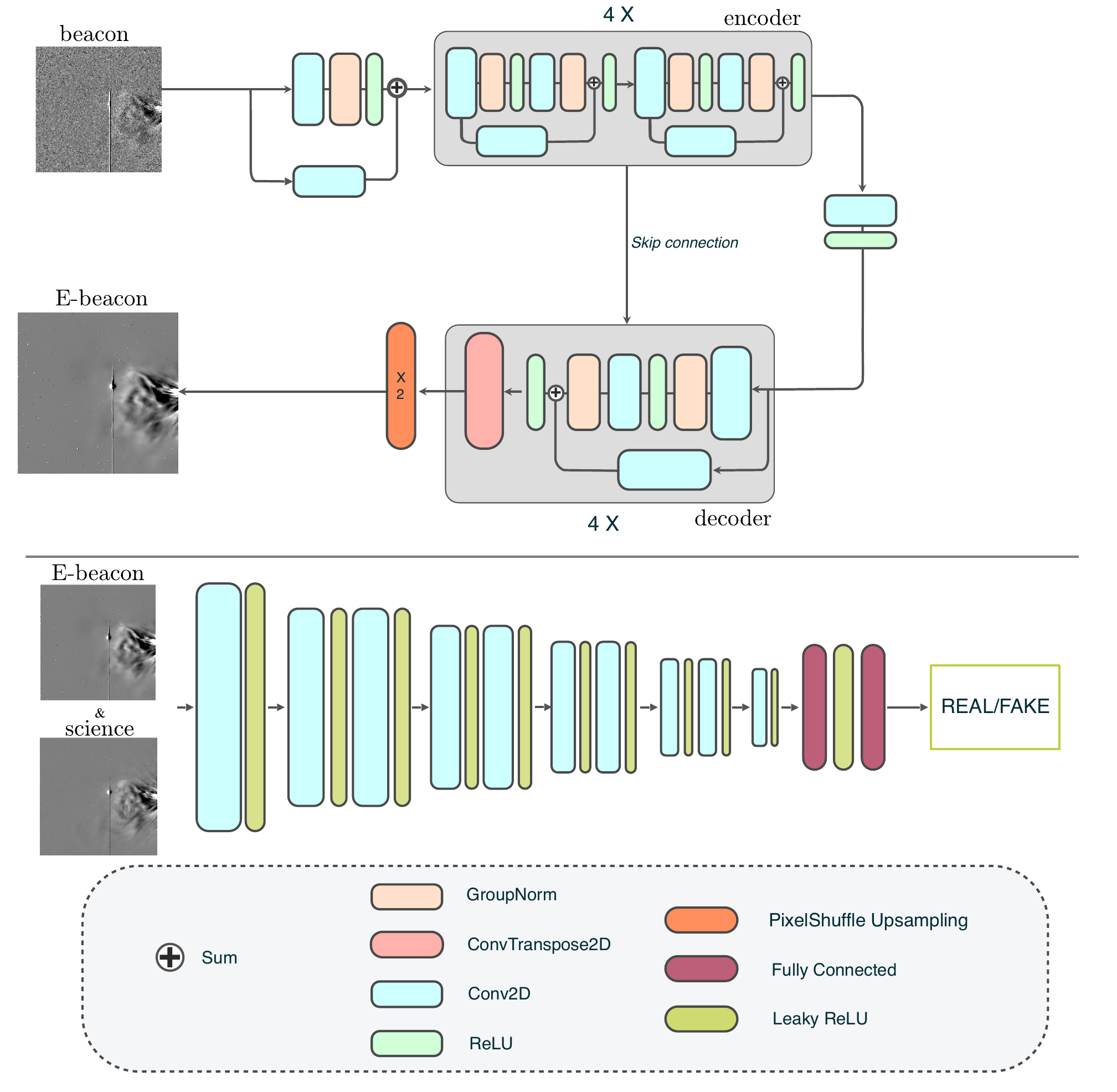}
    \caption{Schematic overview of the first network from our pipeline. In the top panel, the generator network architecture is based on ResUNet, with PixelShuffle layers for super-resolution. In the bottom panel, the discriminator is a small convolutional neural network predicting images to be real or generated. We use respective feature space sizes of [64, 128, 256, 512] for the four residual blocks of the generator.}
    \label{fig:NN1}
\end{figure}

Figure~\ref{fig:NN1} shows the architecture of both the generator and the discriminator. This type of diagram helps visualize how complex models are structured and how different parts of the model work together.
The generator, shown in the upper panel, is based on a ResUNet structure~\cite{zhang2018road}, a type of neural network often used for tasks like image denoising~\cite{lei2023ct}.

It processes the input by passing it through a sequence of functional layers. Each colored block in Figure~\ref{fig:NN1} represents a different functional layer that transforms the input image step-by-step, gradually extracting important features. The connections between the blocks show how data flows through the network, from the initial input to the final output. The two collections of blocks outlined in gray represent groups of layers that the input passes through multiple times.

As the data passes through the encoder, the feature space expands from 64 to 512 dimensions while the spatial resolution shrinks. The decoder then reverses this process, increasing spatial resolution and reducing the feature dimensions. At the output, the generator uses PixelShuffle to expand the spatial resolution by a factor of $X$, producing a single-channel, i.e. grayscale, image (the E-beacon frame).

The discriminator, shown in the lower panel, uses both science and E-beacon images as input and passes them through a series of layers to extract distinctive features. At the end, the features are aggregated and the model uses a fully connected layer (shown in red) to classify the input images as either "real" or "generated".

During training, both networks are optimized using a combination of loss functions. The loss function is used to guide the model's training by evaluating how well a predicted output matches the ground truth data. Models are trained to minimize the loss function. The generator is trained to minimize:

\begin{equation}\label{eq:loss}
    L_{G} = \lambda_{1} L_{c} + \lambda_{2} L_{\mathrm{G}}^{\mathrm{gan}},
 \end{equation}
 
with $L_{c}$ the Mix loss introduced in~\cite{zhao2016loss}, which combines the MSSIM (Multi-scale SSIM) loss~\cite{wang2004image} and $l1$ loss functions. $l1$ loss computes the pixel-wise absolute error between prediction and ground truth images, while the MSSIM loss compares how structurally similar the two images are to each other. Combining both losses allows us to leverage their individual strengths, with the $l1$ loss being more focused on fine details while the MSSIM loss helps to preserve overall structure.

The term $L_{\mathrm{G}}^{\mathrm{gan}}$ is based on the Binary Cross Entropy (BCE) loss function. This loss uses the output of the discriminator network. Including this term in the loss function of the generator ensures that its own training is linked to that of the discriminator. The BCE loss is designed to assess how accurately the model can predict a binary outcome, in our case, whether an image is real or generated. It is given by:

\begin{equation}
BCE(x_n,y_n) =  -1 [y_n  \cdot \log (x_n) + (1-y_n) \cdot  \log(1-x_n].
\end{equation}

By applying a sigmoid function to the output, we obtain a probability between 0 (the image is definitely generated) and 1 (the image is definitely real) for each input image from our discriminator, with the sum of the probabilities amounting to 1. We employ an approach first introduced by \citeA{jolicoeur-martineau2018}, and define $L_{\mathrm{G}}^{\mathrm{gan}}$ as:

\begin{equation}
    L_{\mathrm{G}}^{\mathrm{gan}} = BCE\left( D\left( \mathrm{pred} \right) - D\left( \mathrm{GT} \right), 1 \right),
\end{equation}

with $D(pred)$ being the predicted probability of the E-beacon image, and $D(GT)$ being the predicted probability of the science image (GT). For our generator’s loss function, we want the difference between the $D(pred)$ and $D(GT)$ to be as close to 1 as possible, meaning that the E-beacon image is so convincing that the discriminator falsely classifies it as a science image.

We use weights $\lambda_{1} = 2$ and $\lambda_{2} = 5\times10^{-4}$ for Equation~\ref{eq:loss} to fine-tune the relative importance of the individual loss functions.

The loss function used to train the discriminator network similarly follows the approach in \citeA{jolicoeur-martineau2018} and is defined as follows:

 \begin{equation}
     L_{\mathrm{D}}^{\mathrm{gan}} = \frac{1}{2}   BCE\left( D\left( \mathrm{GT} \right) - D\left( \mathrm{pred} \right), 1 \right) + \frac{1}{2}   BCE\left( D\left( \mathrm{pred} \right) - D\left( \mathrm{GT} \right), 0 \right),
 \end{equation}
with $D$ being the discriminator, GT the real science data, pred the output of the generator (E-beacon).

We use a super-resolution factor $X = 2$ due to computing limitations and to improve training time. To ensure stable training, we introduce a short “warmup” phase lasting four training passes (epochs), during which the contribution of the adversarial loss term ($L_{\mathrm{G}}^{\mathrm{gan}}$) is temporarily disabled by setting $\lambda_{2}=0$. This helps the generator start learning meaningful structure before the discriminator starts applying pressure to improve accuracy.

To train the networks, we use the Adam optimizer~\cite{adam}, which automatically adjusts how the model updates itself to improve learning efficiency. The learning rate (which controls the step size with which the optimizer moves towards the minimum of the loss function) is set to $10^{-5}$ for the generator and $10^{-6}$ for the discriminator.
We reserve $10\%$ of the dataset as a validation set. This portion is not used to update the model parameters but helps monitor progress and prevent overfitting, which happens if the model exhibits strong performance on the training data, but poor performance on new, unseen data. Training is stopped early if the model’s performance on the validation set stops improving.

\subsection{Increasing Temporal Resolution}\label{seq:NN2}

Besides the low resolution and high noise, the low cadence of the beacon mode makes tracking CMEs across several frames difficult for fast and interacting events. Furthermore, it reduces the readability of J-maps compared to one generated from science data, with tracks being more challenging to separate. 

\begin{figure}[ht!]
    \centering
    \includegraphics[width=\textwidth]{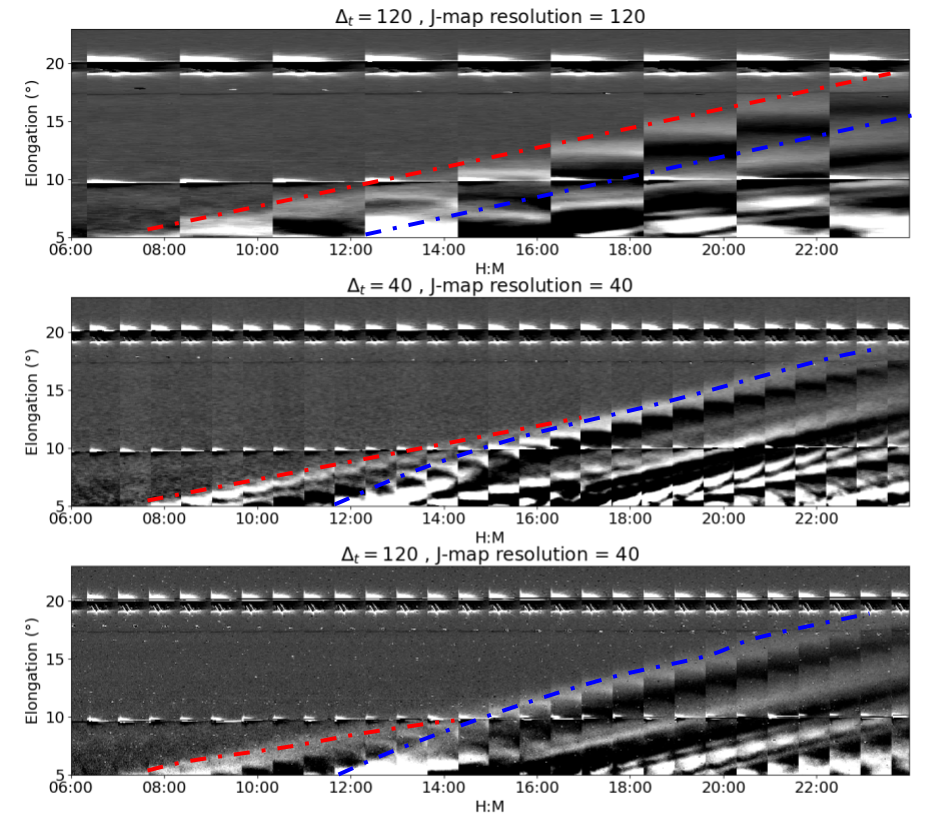}
    \caption{Science J-maps at varying time-axis resolutions, using varying $\Delta_t$ for creating the running difference images. Higher time-axis resolution J-maps improve the visibility and separation of CME tracks even at higher $\Delta_t$}
    \label{fig:JPLOTScomparison}
\end{figure}

We create J-maps using science data with time-axis resolutions of 40 and 120 minutes, meaning that the data is spaced either 40 or 120 minutes apart along the time-axis. Additionally, we vary $\Delta_t$, the time interval between subsequent images used to create running differences. We choose $\Delta_t$ = 40 minutes and $\Delta_t$ = 120 minutes. Figure~\ref{fig:JPLOTScomparison} shows how varying these two parameters impacts the visual quality of J-maps, and consequently the tracking of CMEs. The top-most panel shows a J-map with a time-axis resolution of 120 minutes, created using a $\Delta_t$ of 120 minutes for the running difference images. Effectively here, two out of every three images are completely discarded representing what would be observed if the science data were downlinked at the cadence of beacon mode data. Two distinct CMEs appear to be visible in the J-map, one tracked in blue, the other one in red. The CME associated with the blue track appears to closely follow the one associated with the red track, with seemingly no interaction between the two. The middle panel shows a J-map with a time-axis resolution of 40 minutes, created using a $\Delta_t$ of 40 minutes for the running difference images. In this plot, a different picture seems to emerge, with the blue CME now appearing to overtake the red CME, instead of following behind it. The bottom panel shows a J-map with a time-axis resolution of 40 minutes, created using a $\Delta_t$ of 120 minutes for the running difference images. The evolution of both CMEs agrees with that of the middle panel---the CMEs merge after the blue CME overtakes the red one. This highlights the importance of using high time-axis resolution for J-maps to enable accurate tracking of CMEs. Even at a higher $\Delta_t$ of 120 minutes, the respective CMEs’ trajectories are easily identified if the time-axis resolution of the J-map is set to 40 minutes. In contrast, setting the time-axis resolution to 120 minutes for the same $\Delta_t$ produces a trajectory that is markedly different from that seen in the higher time-axis resolution J-maps.

Beacon images are available at a cadence of 120 minutes, consequently limiting the $\Delta_t$ for creating running difference images as well as the time-axis resolution of J-maps to 120 minutes. While creating running difference images at a $\Delta_t$ smaller than 120 minutes is not possible for beacon data, it is possible to interpolate between running difference images to create a J-map with a time-axis resolution of 40 minutes, while keeping $\Delta_t$ at 120 minutes, as shown in the bottom panel of Figure~\ref{fig:JPLOTScomparison}. To accomplish this, we propose training a model using science running difference images created at a $\Delta_t$ of 120 minutes (to match the $\Delta_t$ of beacon data), with a time-axis resolution of 40 minutes. The model then learns to create IE-beacon data at the standard $\Delta_t$ of 120 minutes, but a time-axis resolution of 40 minutes. This should enhance the visibility of individual CMEs in J-maps compared to using $\Delta_t$ of 120 minutes and time resolution of 120 minutes, thus facilitating tracking.

We use the method described by \citeA{huang2022real}, called RIFE (Real-Time Intermediate Flow Estimation for Video Frame Interpolation). The model introduced in RIFE is an efficient deep-learning frame interpolation method tailored to generate an intermediate image $I_{t}$ from a pair of images $I_{0}$ and $I_{1}$, primarily for increasing video frame rate. 
To this extent, the direction and magnitude of movement of the pixels, known as optical-flow, between the two frames $I_{0}$ and $I_{1}$ is predicted. Knowledge of the optical-flow can then be used to predict what the image $I_{t}^{pred}$ at the intermediate target timestep $t$ looks like. During training, the model learns to minimize a loss function expressed as follows:
\begin{equation}
      L  = L_{\mathrm{rec}} + \lambda L_{\mathrm{dis}}
      \label{eq:RIFEloss}
\end{equation}

$L_{\mathrm{rec}}$ is the $l1$ loss (absolute difference) between the interpolated image $I_{t}^{pred}$ and a ground truth (i.e. science) image at the same timestep $I_{t}^{gt}$. $L_{\mathrm{dis}}$ gives the $l2$ loss (squared difference) between the predicted optical-flow, computed using $I_{0}$ and $I_{1}$, and the true optical flow, computed using $I_{t}^{gt}$. We set $\lambda=0.0001$ to assign weights to the individual loss terms. Similarly to RIFE, we use the AdamW~\cite{adamw} optimizer, with a learning rate of $10^{-5}$ and a weight decay of $10\times10^{-3}$. Utilizing a weight decay parameter can help prevent overfitting by adding a penalty term to the loss function, preventing it from growing too large, and effectively improving generalizability and stability.

\subsection{Combining Tracking in Images and J-maps}\label{sec:tool}

\begin{figure}
    \centering
    \includegraphics[width=\linewidth]{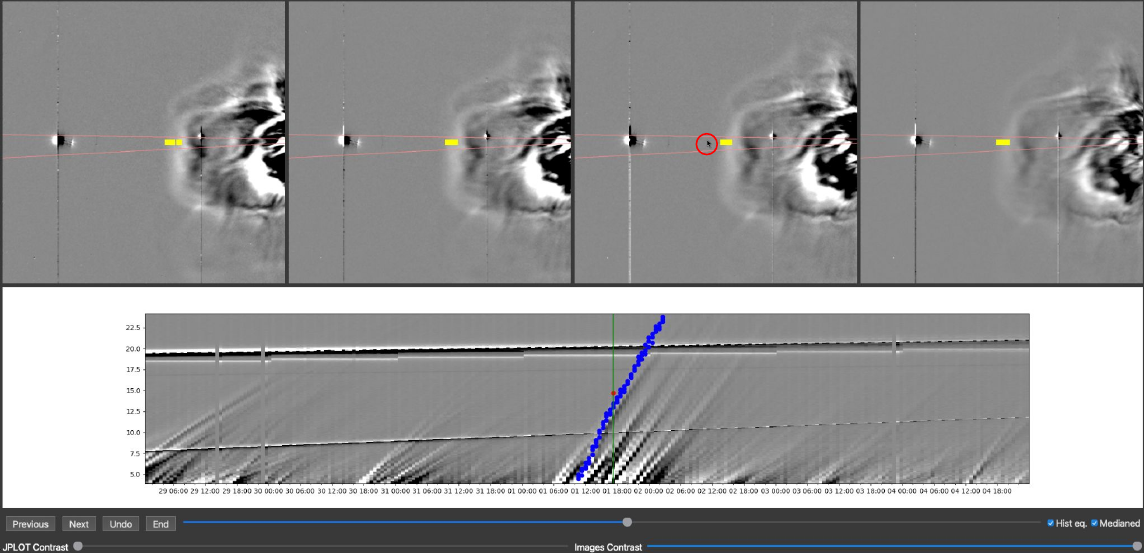}
    \caption{The proposed interface for CME tracking simultaneously in images and J-maps. We display the J-map for the event and four running differences at a time. A vertical green line indicates the third image position on the J-map, and a red dot indicates the mouse position (highlighted with a red circle in the third image). The tracked points are displayed as yellow squares on the images and blue dots on the J-map. The button allows you to change the images displayed, undo the last annotated point, and end the tracking by saving it. The topmost slider allows one to navigate through the J-map timeline, and the bottom two sliders change the contrast of the J-map and images for better visibility of different features at different elongations. }
    \label{fig:interfaceannotation}
\end{figure}

The tracking of CMEs in the HI field of view is frequently done using J-maps. However, this representation loses global spatial context for the different structures of the CME and makes it difficult to understand CME interactions with other CMEs and detangle the different tracks to follow. To circumvent this challenge and improve tracking efficiency, we introduce a new tracking tool combining sequences of HI running differences images and the corresponding J-map for specific events. We show the tool interface in Figure~\ref{fig:interfaceannotation}. Buttons and sliders allow for easy navigation between frames and J-map, and tracking is done directly on the images, by selecting the points of interest to track. We also introduce two sliders to control the contrast of the images and the J-map to increase the visibility of CMEs as they get fainter in the second half of the HI image. Finally, the displayed images are indicated with vertical green lines in the J-map, and the mouse position is marked with a red dot to correlate J-map tracks and image features visually. We show the ecliptic plane in each image to know where the features of interest should be tracked.

When tracking a CME in the images with the help of a J-map, we select multiple points around the perceived leading edge of the CME front to indicate uncertainty, as following the same feature is impossible on an image, even more so on noisy beacon data.

\section{Results}

To assess the performance and utility of our machine learning method, we separate the evaluation into three parts. First, we evaluate the first neural network output, which we call E-beacon, and assess the improvement in the noise level and fidelity to science data for our test events. We then focus on the output of the second neural network, namely IE-beacon created from E-beacon to match the science time resolution of $40$ minutes. Finally, we evaluate the different data types from a tracking perspective using science data as a baseline for CME tracks.

To evaluate the generated running difference images signal-to-noise-ratio (SNR) and the SNR improvement over beacon images, we measure the peak signal-to-noise ratio (PSNR), the structural similarity \cite<SSIM;>{wang2004image}, and the Mean Square Error (MSE) between beacon, E-beacon, IE-beacon, and science images. PSNR and SSIM indicate the overal fidelity to science data, while the MSE reflects a reduction in noise and distortion.
We define the MSE as:
\begin{equation}
    \mathrm{MSE}(K) = \frac{1}{n n} \sum^{n-1}_{i=0}\sum^{n-1}_{j=0}[I_{s}(i,j) - K(i,j)]^2,
\end{equation}
with K a beacon, E-beacon or IE-beacon image and $I_{s}$ the corresponding science frame. n is the size of the image in pixels, i and j the row/column of each pixels. PSNR is defined as :
\begin{equation}
    \mathrm{PSNR}(K) = 20 \cdot log_{10}(R). - 10 \cdot log_{10}(\mathrm{MSE}(K)),
\end{equation}
with R the data range of the noisy image (ie. beacon, E-beacon, IE-beacon).

The Structural Similarity Index (SSIM) is a perceptual metric that quantifies image similarity by jointly evaluating luminance, contrast, and structural information. For two image patches xx and yy, SSIM is defined as:
\begin{equation}
\mathrm{SSIM}(K, I_s) = \frac{(2\mu_x\mu_y + C_1)(2\sigma_{xy} + C_2)}{(\mu_x^2 + \mu_y^2 + C_1)(\sigma_x^2 + \sigma_y^2 + C_2)}
\end{equation}

where $\mu_x$, $\mu_y$ are the local means, $\sigma_{x}^2$, $\sigma_{y}^2$ are the variances, $\sigma_{xy}$  is the covariance, and $C1 = (0.01R)^2$, $C2 = (0.03R)^2$ are stabilizing constants.

As we want to evaluate the perceptual improvement over beacon data (because of their low contrast after normalization, running difference images need to have their contrast increased for users to visualize them), we pre-process the running differences as follow:
\begin{equation}
    I =  I  c + (-\frac{c}{2} + 0.5),
\end{equation}
with $c$ the contrast value and $I$ a running difference. We select $c=5$ through experimentation, to maximize faint CME fronts visibility.

\subsection{Enhancement Evaluation}
    We evaluate the performances of the first neural network used for super-resolution and denoising described in Section~\ref{seq:NN1}. To this extent, we compare an upsampled version of the beacon data using bicubic upsampling to the corresponding science images and similarly compare E-beacon to science images. As we trained neural network 1 to produce an upsampled image of only twice beacon resolution, resulting in a 512 pixels $\times$ 512 pixels, we downsampled science data and upsample beacon using Bi-cubic spline interpolation to the 512 pixels $\times$ 512 pixels resolution for all comparisons.
    
    We show the results for this comparison in Table~\ref{tab:enhanced}. 
    The raw beacon data exhibits a relatively low PSNR ($19.04$) and SSIM ($42.40 \times10^{-2}$), with a high MSE ($33.84\times10^{-3}$), which confirms the significant noise increase and quality reduction when compared to the science data.
    After the first neural network processing, the E-beacon data shows a notable improvement in all three metrics. PSNR increase to $26.10$ attests to a substantial noise reduction, while decreased MSE confirms a lower pixel-wise error. We also see a rise in SSIM to $67.70\times10^{-2}$, indicating better structural similarity to science data.
    The IE-beacon data similar performance to E-Beacon across all metrics, with a PSNR of $25.81$ and with an SSIM of $67.28\times10^{-2}$. This shows that the interpolated images, are faithful to the science data and the features motion correclty infered.
    
    These results demonstrate substantial improvement in the quality of beacon data. While the first network provides significant enhancement with increased PSNR and SSIM and a decreased MSE, our second neural network interpolates E-beacon faithfully to science. The presented performances confirm the effectiveness of this approach for restoring and enhancing low-resolution beacon data while keeping important information. Furthermore, our second neural network increases the temporal resolution while keeping similar performance to E-beacon data.

       \begin{table}[!h]
        \caption{Error metrics of the test set comparing beacon data, the E-beacon data, and the IE-beacon to science data. }
        \label{tab:enhanced}
            \centering
            \begin{tabular}{l c c c}
                \hline
                 Data type  & PSNR ($\uparrow$) & MSE($\downarrow$) & SSIM($\uparrow$)  \\
                \hline
                  beacon  & $19.04$  & $33.84\times10^{-3}$ & $42.40 \times10^{-2}$\\
                  E-beacon  & $26.10$  & $5.15\times10^{-3}$ & $67.70\times10^{-2}$ \\
                  IE-beacon & $25.81$  & $5.39\times10^{-3}$ & $67.28\times10^{-2}$ \\
               \hline
            \end{tabular}
    \end{table}

\begin{figure} [ht!]
       \centering
       \includegraphics[width=\textwidth]{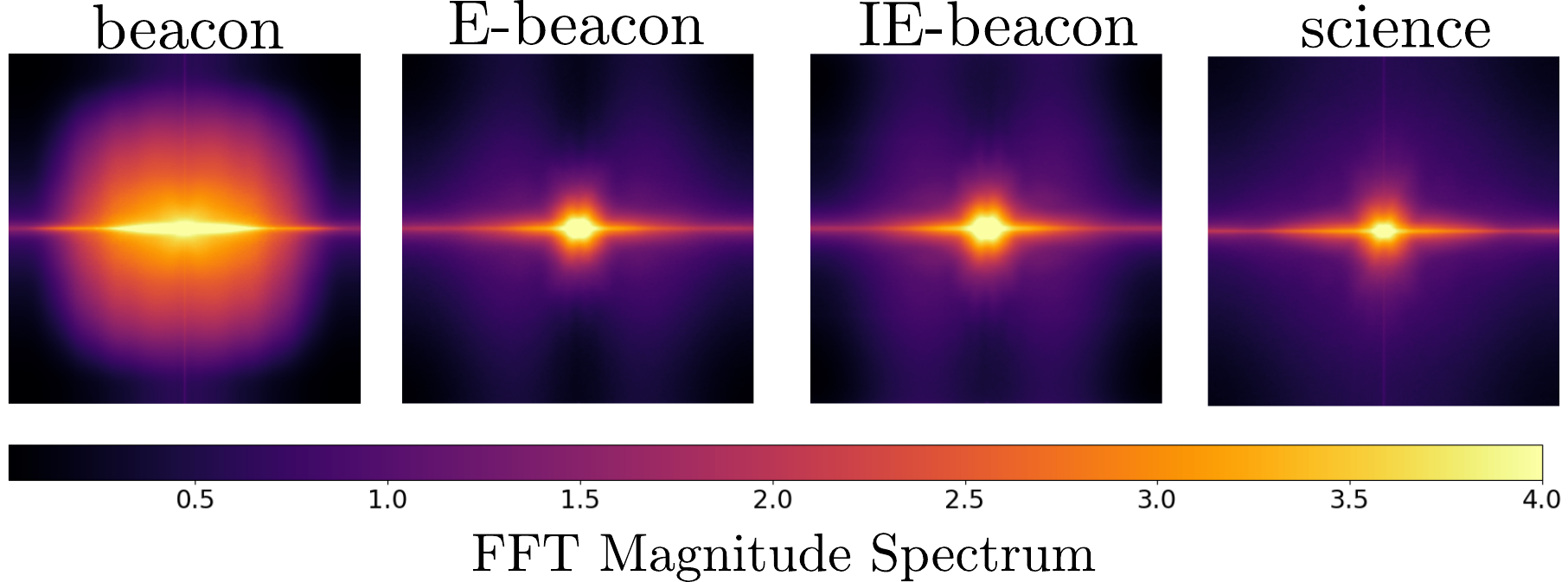} 
       \caption{E-beacon and IE-beacon improvement visualized using the average FFT spectrum of the test set for beacon, E-beacon, IE-beacon and science images. The pattern of beacon differs greatly from the science with high-frequency noise. E-beacon and IE-beacon spectrum shows greater similitude to science images.}
       \label{fig:NN1improvement}
    \end{figure}

    To further analyze the differences in image SNR and structure, we compute the averaged Fast Fourier Transform (FFT) spectrum of the images in the beacon, E-beacon, IE-beacon and science data test sets. The FFT spectrum provides insight into the spatial frequency distribution, highlighting noise characteristics and structural fidelity. Figure~\ref{fig:NN1improvement} shows the spectrum for all four data types. The X and Y axes correspond to spatial frequencies repetition in x and y direction in cycles/pixels. The center represents low spatial frequencies (large-scale structures), and higher frequencies closer to the edges correspond to finer details/noise.   
    The FFT spectrum of the raw beacon data exhibits a bright and uniform intensity across a large band of frequencies corresponding to significant high-frequency noise. In contrast, the E-beacon and IE-beacon images appear more similar to the pattern displayed in the science image, with brightness concentrated near the center of the spectrum. This similarity in pattern shows that the first neural network effectively suppresses high-frequency noise while preserving the structure and characteristics of the original science data, and the second neural network preserves these characteristics. This similarity further supports the effectiveness of the denoising and upsampling process by recovering relevant image features while reducing noise-induced artifacts introduced by compression and binning.

\begin{figure}
 \centering
  \includegraphics[width=0.9\textwidth]{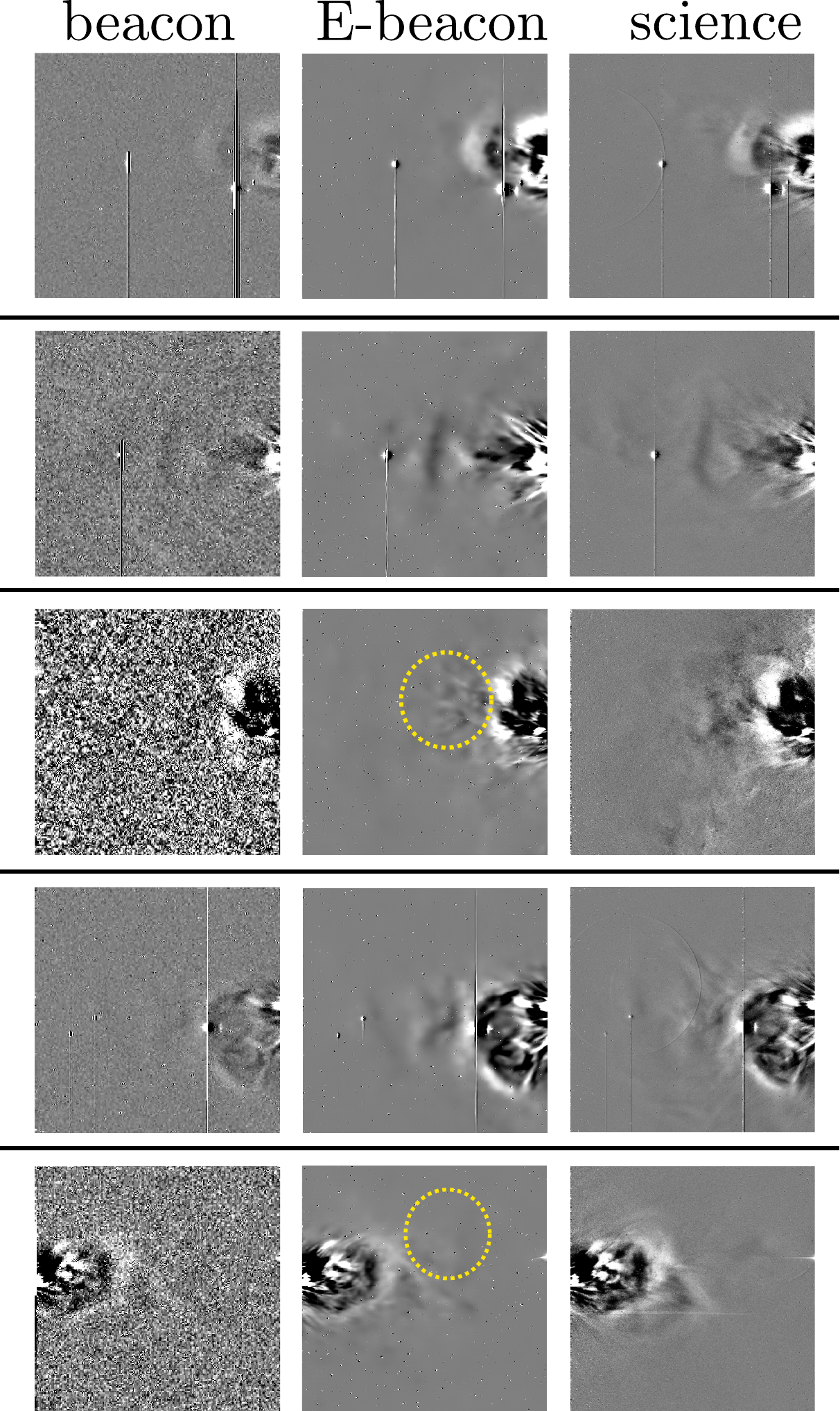}
\caption{Comparison between beacon, E-beacon and science running difference at the 120 minutes $\Delta_t$. The images correspond to CME1, CME3, CME9, CME14 and CME33 from our test set. For each image, the contrast is chosen for the best visibility of the CME front.}
\label{fig:imgnn1}
\end{figure}

    Figure~\ref{fig:imgnn1} presents qualitative examples of the improvements in CME visibility achieved through our denoising method, comparing beacon, E-beacon, and science images. The beacon images exhibit high noise levels, significantly hindering the visibility of CME fronts, especially beyond the first quarter of the image. The presence of bright planets within the field of view further exacerbates this issue, blending CME features with the noisy background and making their identification more challenging.
    In contrast, the E-beacon images recover most of the CME structures, improving their contrast and overall visibility. The CME fronts appear more continuous and discernible, allowing for better tracking of key feature propagation.
    However, faint background features are disproportionately amplified in some of the E-beacon images (highlighted by yellow circles). This enhancement leads to cloud-like patterns appearing more prominent than expected, potentially altering the perceived intensity distribution of CME features. Furthermore, the E-beacon result produces over-smoothed features for noisier beacon images while trying to recover structures from the overall noise. This tendency to overemphasize low-intensity structures and reduce the sharpness of key features suggests an area for further refinement in the enhancement algorithm.

\subsection{Increasing Time Resolution}

\begin{figure}
  \centering
  \includegraphics[width=\linewidth]{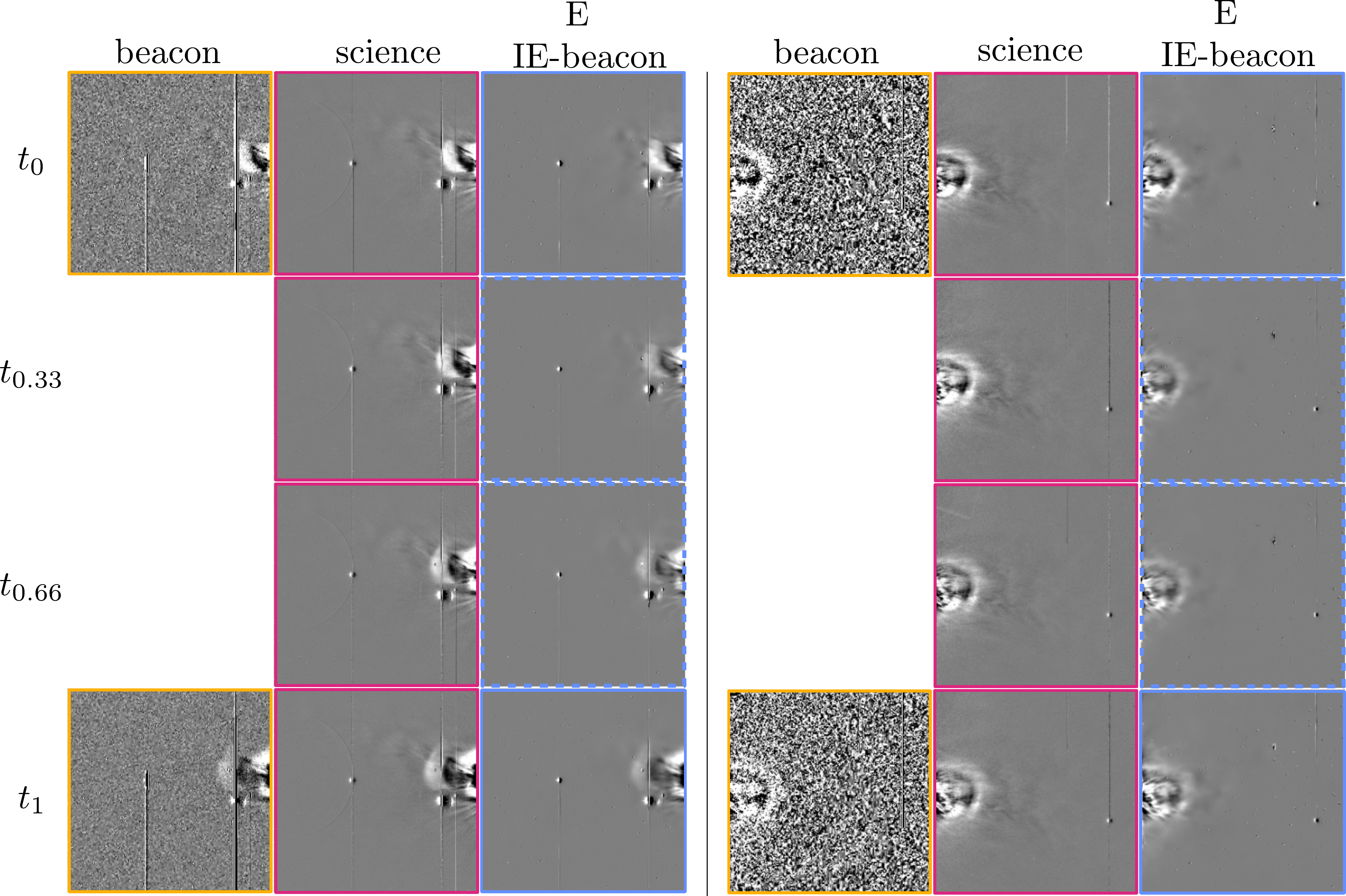}
\caption{Beacon (yellow), science (pink), E-beacon (blue) and IE-beacon (dashed blue) sequences part of CME1 and CME38 (left and right). We show the original beacon and E-beacon frames at $t_{0}$ and $t_{1}$, and the interpolated IE-beacon at $t_{0.33}$ and $t_{0.66}$. The time resolution of beacon (120 minutes), is increased to 40 minutes with IE-beacon data.}
\label{fig:imginterpolated}
\end{figure}

Figure~\ref{fig:imginterpolated} presents qualitative results of the interpolation method, showing two pairs of consecutive beacon images, their E-beacon versions, the two IE-beacon frames, and science frames for the whole sequence as comparison. The generated intermediate images exhibit smooth transitions between the two input frames, demonstrating temporal consistency in CME evolution. The key structural features of the CME, including its bright front and core, are well-preserved, while the propagation of the front remains coherent across interpolated frames.
Notably, the interpolated images maintain the CME's overall shape and intensity distribution, avoiding the introduction of significant artifacts or distortions. The motion of the front visually aligns well with the expected temporal evolution, suggesting that the interpolation method effectively reconstructs intermediate states while preserving the physical integrity of the CME structures.

\subsection{Evaluating of CME Front Tracking}

\begin{figure}[ht!]
    \centering
    \includegraphics[width=\textwidth]{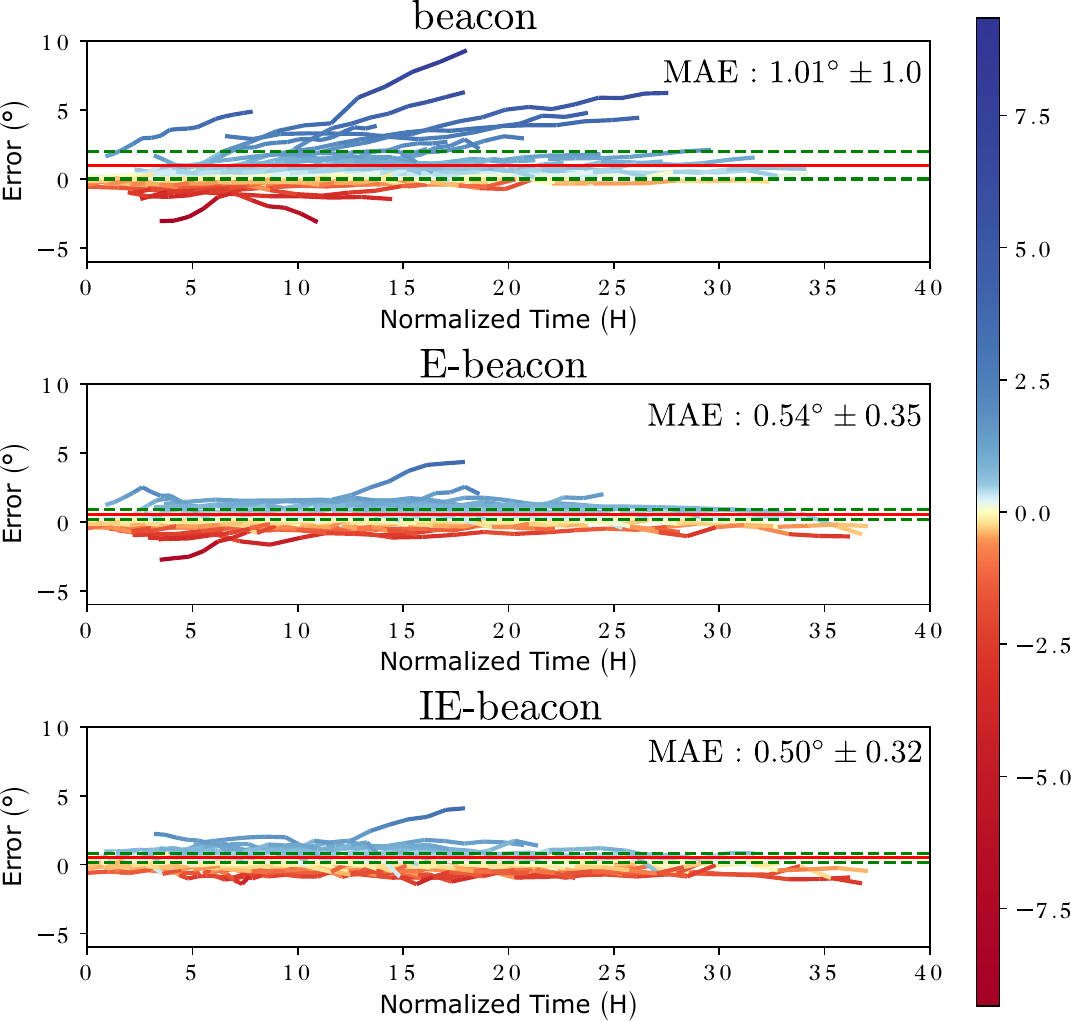}
    \caption{Comparison of track error for each CME. We normalized the time on the X-axis by the earliest point between all data types for each event. The color indicates the elongation error compared to tracks derived from science data.}
    \label{fig:trackserror}
\end{figure}

To understand the effect of data quality and noise on tracking CME features used for forecasting, we use the tracking tool introduced in Section \ref{sec:tool} to track the CME's front for our test set. To reduce bias and simulate real-time conditions, we first track each CME in beacon data, followed by E-beacon, IE-beacon, and science data in order. For each CME, we used the approximate appearance time in the HI1 field of view as the starting point and tracked the front as best as possible until it became too faint to distinguish it from the background. As we want to avoid another layer of uncertainty brought by fitting algorithms such as FPF \cite{rouillard2008first,sheeley1999continuous}, HMF\cite{lugaz2009deriving,lugaz2010accuracy} or more advanced drag-based models such as ELEvoHI \cite{rol16}, we use the science tracks as a reference and compare the other data type tracks to them. In doing so, we do not evaluate the difference in terms of arrival time, just the deviation from science tracks.
We show the deviation from science for each track in Figure \ref{fig:trackserror}. Beacon tracks, on average, do not deviate largely from science tracks, with a MAE of $\sim 1^\circ$ elongation. However, some CME onsets were more challenging to identify, and 9 from 48 diverged strongly due to a fainter front or were difficult to follow due to higher noise levels. E-beacon tracks with an MAE of 0.54 degrees show less deviation at higher elongation. IE-beacon further improves tracking (MAE of $0.50^\circ$ elongation) with less divergence at the beginning of the tracks and better precision at higher elongation.
 
\begin{figure}[h!]
    \centering
    \includegraphics[width=\textwidth]{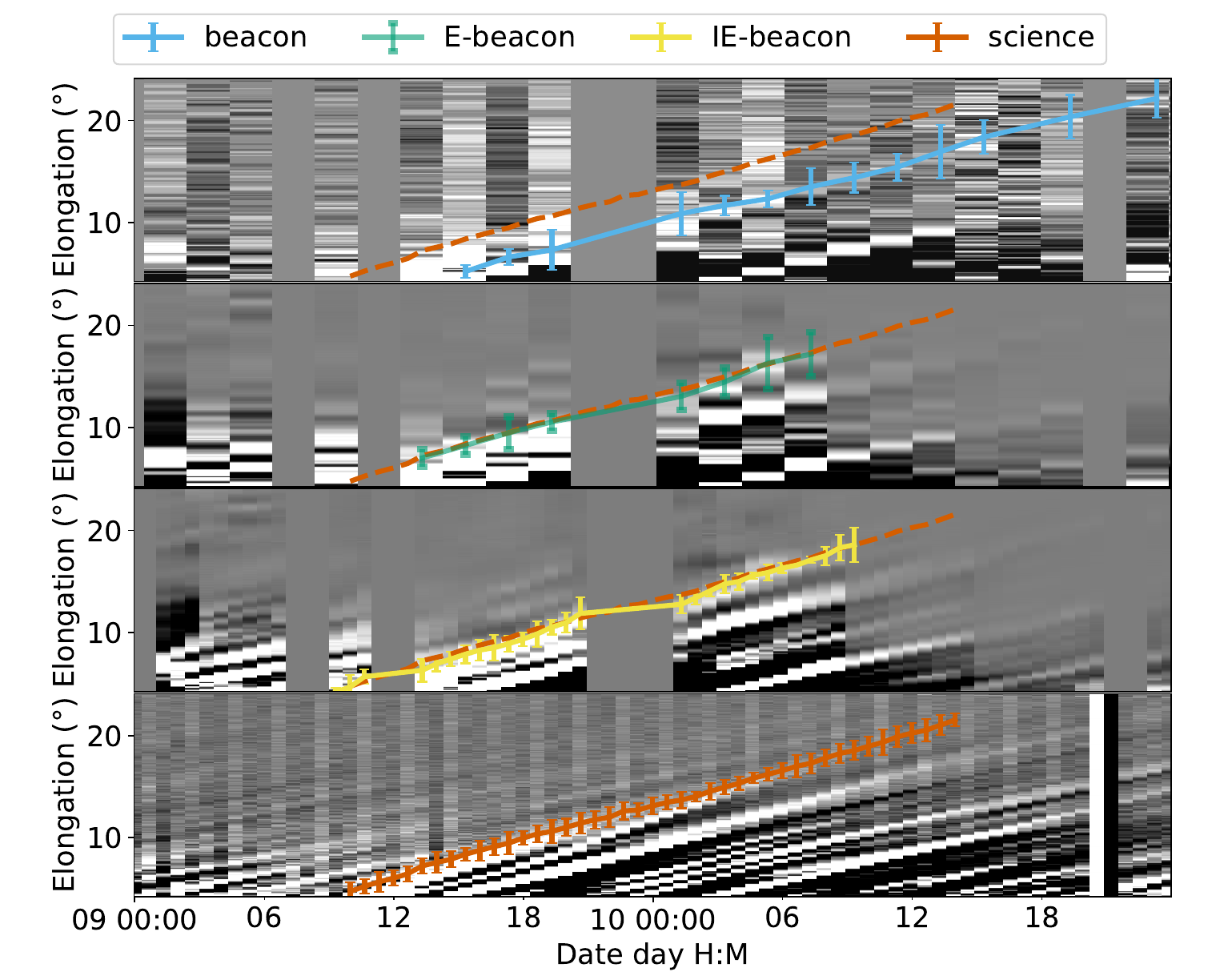}
    \caption{Comparison of J-maps for in order beacon, E-beacon, IE-beacon and science. For each J-map, we show the track obtained for the corresponding data type and the science track. We show the uncertainty from tracking with five times the standard deviation for easier visualization.}
    \label{fig:jplotscomp}
\end{figure}

Figure~\ref{fig:jplotscomp} presents the CME front tracking results for three examples, comparing beacon, E-beacon, and IE-beacon data. For each case, the extracted tracks are shown alongside the science reference (dashed line), with the measurement points and their associated uncertainties marked on. The raw beacon data exhibits the highest uncertainty, with more significant variations in front positions and occasional deviations from the expected trajectory. Such error is primarily due to the high noise levels, which reduce the CME feature's visibility and makes it challenging to consistently identify the front across frames.

In contrast, the E-beacon data significantly improves the visibility of CME structures, leading to more stable and coherent tracks that better align with the science reference. The uncertainties are notably reduced, indicating greater confidence in the extracted positions. The IE-beacon data further refines the tracking by enhancing temporal consistency, making it easier to follow CME features between consecutive images. This results in smoother tracks, further improving the accuracy and continuity of front detection. Overall, the improvements in visibility and feature continuity demonstrate the effectiveness of our method in reducing noise and enhancing CME tracking.

However, one remaining challenge is the presence of data gaps in beacon data, which makes it harder to continue tracking CMEs with precision. Due to the scale of several hours for most of them, interpolation of images is impossible.

\section{Discussion}

Our results demonstrate that E-beacon data significantly improves CME visibility compared to standard beacon images, reducing noise and enabling enhanced contrast. This enhancement leads to more stable and accurate CME tracking. Furthermore, introducing interpolation in time further refines feature continuity, facilitating smoother front propagation across frames. Combining these improvements leads to reduced tracking uncertainty and better alignment with science data, offering a more reliable method for CME analysis with STEREO-A/HI1 data in near-real-time. While we do not assess the accuracy of an actual arrival time prediction, CME fronts tracked in E-beacon and IE-beacon are more similar to CME fronts tracked in science data compared to beacon data.

However, the enhanced data produced still suffers from artifacts when trying to increase the brightness of areas more than it should and struggles to recover some details of the CME structure. More work should be done to recover small structures from the noise and reduce the over-smoothing of faint areas. Adding temporal context to the first neural network (processing sequences of images instead of single images) should help improve the structures and appearance of CME, reducing the impact of the noise by recovering information over multiple frames. As such temporal aspects require 3D convolutions (to process the extra dimension added by sequences of images), it would be important to balance out the network size to keep training and memory consumption reasonable. 

The critical challenge with beacon data remains the larger data gaps in the downlinked data that are impossible to interpolate over, with only a small time frame of the CMEs propagation visible. As the gaps are not systematically consistent across the different instrument fields of view, adding STEREO/COR2 as part of the overall pipeline could mitigate the gaps. In future work, we aim to fine-tune the two neural networks on STEREO/HI2 and COR2 data to produce enhanced data from the three fields of view of the STEREO mission. Furthermore, extrapolating each field of view into the next one (COR2 features propagated to HI1) would allow better tracking and understanding of CME dynamics.

Finally, we used running difference images as input to improve the visibility of CMEs and other transients and allow better and more stable training of the different neural networks. Improving the pre-processing of Level 2 images to enhance the visibility of CMEs and adapting our method for such data would allow us to interpolate L2 images directly and obtain running differences with improved details and structure appearance of CMEs in running differences with lower $\Delta_t$.  

Addressing these limitations and adding these improvements to the methods presented in this paper is considered future work.

\section{Conclusion}

In this paper, we proposed enhancing STEREO-A/HI1 near-real-time data quality (through reducing noise and increasing spatial resolution) and increasing temporal resolution to mitigate the effect of the heavy compression and binning applied onboard due to the limited broadcast bandwidth. We applied a first neural network to reduce the noise and increase the resolution of beacon running differences. The new data called E-beacon greatly improves the signal-to-noise ratio over beacon data while keeping important information with a PSNR of $26.10$ and SSIM of $67.70\times10^{-2}$. These results prove that it is possible to recover meaningful information from beacon data and improve the visibility of CMEs in near-real-time despite the low bandwidth rate of the beacon mode.

We then trained a state-of-the-art frame interpolation method to increase the temporal resolution of the newly enhanced data. 
The new data, called IE-beacon, produces intermediary enhanced running differences images (with a SSIM of $67.28\times10^{-2}$, showing similar performances to E-beacon) and facilitates the tracking of features within the images due to smoother and smaller movement.  This improvement is evident within J-maps with smoother tracks and a better understanding of the different bright parcels. During inference (execution of the trained models), on an NVIDIA RTX 4090 GPU it takes 2ms to produce an E-beacon frame, and 3ms for an IE-beacon frame (comparatively, on CPU both models take 0.3s). These very fast processing times would easily allow us to deploy our method and provide E/IE-beacon data for real-time forecasting applications.

Finally, we analyzed the potential for tracking improvement from the two new data types proposed in this paper. We proposed a new tracking tool that we use to track 48 CMEs for four different datasets and use science data tracks as a reference. Both E-beacon and IE-beacon bring significant improvement in tracking, with tracks closer to science, smoother tracks, and fewer large deviations at higher elongation due to low visibility and cadence.

The work presented in this paper doesn’t need to be limited to STEREO/HI1 data but paves the way for its application to other missions and data types. By selecting test events observed from the L4 and L5 points, we introduce an additional tool for potential use on future missions, such as Vigil, if, for example, bandwidth needs to be reduced, but real-time tracking of CME is needed.

%
%

\section*{Data Availability Statement}\label{sec:code}

The data files used in this paper can be downloaded from \url{https://stereo-ssc.nascom.nasa.gov/pub/} (or using the code provided). We provide the code for processing the data (downloading, reduction...), models, tracking tool,  training and testing code under \citeA{lelouedec2025zenodo}. The models' weight and CME list can be found at \citeA{lelouedec2025}.
HELCATS WP2 and WP3 catalogs can be found respectively at \url{https://www.helcats-fp7.eu/catalogues/wp2_cat.html} and \url{https://www.helcats-fp7.eu/catalogues/wp3_cat.html}.
This research used SunPy \cite{sunpy_community2020}, AstroPy \cite{astropy2022} and Pytorch \cite{paszke2019pytorch}.

\acknowledgments
This research was funded in whole or in part by the Austrian Science Fund (FWF) [10.55776/P36093]. For open access purposes, the author has applied a CC BY public copyright license to any author-accepted manuscript version arising from this submission. This work is supported by ERC grant (HELIO4CAST, 10.3030/101042188). Funded by the European Union. Views and opinions expressed are however those of the author(s) only and do not necessarily reflect those of the European Union or the European Research Council Executive Agency. Neither the European Union nor the granting authority can be held responsible for them.

\bibliography{references}

\end{document}